\documentclass[12pt]{article}
\usepackage{epsf}

\newdimen\SaveWidth \SaveWidth=\textwidth
\newdimen\SaveHeight \SaveHeight=\textheight
\newdimen\SaveW \SaveW=\textwidth
\newdimen\SaveH \SaveH=\textheight
\advance\SaveWidth by -\textwidth
\advance\SaveHeight by -\textheight
\divide\SaveWidth by 2
\divide\SaveHeight by 2
\advance\hoffset by \SaveWidth
\advance\voffset by \SaveHeight

\catcode`@=11
\def\citenum#1{%
   \expandafter\ifx\csname b@#1\endcsname\relax{\bf ??}\else
   \csname b@#1\endcsname\fi}
\catcode`@=12

\def\slashchar#1{\setbox0=\hbox{$#1$}           
   \dimen0=\wd0                                 
   \setbox1=\hbox{/} \dimen1=\wd1               
   \ifdim\dimen0>\dimen1                        
      \rlap{\hbox to \dimen0{\hfil/\hfil}}      
      #1                                        
   \else                                        
      \rlap{\hbox to \dimen1{\hfil$#1$\hfil}}   
      /                                         
   \fi}                                         %

\def\simge{
    \mathrel{\rlap{\raise 0.511ex
        \hbox{$>$}}{\lower 0.511ex \hbox{$\sim$}}}}
\def\simle{
    \mathrel{\rlap{\raise 0.511ex 
        \hbox{$<$}}{\lower 0.511ex \hbox{$\sim$}}}}

\catcode`@=11
\newdimen\vbigd@men                             

\def\vbig#1#2{{\vbigd@men=#2\divide\vbigd@men by 2%
   \hbox{$\left#1\vbox to \vbigd@men{}\right.\n@space$}}}
\catcode`@=12

\def\etmiss{\slashchar{E}_T}
\def\tg{{\tilde g}}
\def\tq{{\tilde q}}
\def\tchi{{\tilde\chi}}

\def\lsp{{\tilde\chi_1^0}}
\def\tG{{\tilde G}}
\def\ttau{{\tilde\tau}}
\def\tell{{\tilde\ell}}

\def\GeV{{\rm GeV}}
\def\TeV{{\rm TeV}}
\def\Meff{M_{\rm eff}}

\def\fbi{{\rm fb^{-1}}}

\def\mhalf{{m_{1/2}}}

\def\hc{{\rm h.c.}}

\def\dofig#1#2{\epsfxsize=#1\centerline{\epsfbox{#2}}}
\def\dofigs#1#2#3{\centerline{\epsfxsize=#1\epsfbox{#2}%
   \hfil\epsfxsize=#1\epsfbox{#3}}}

\begin{document}

\font\twelvess=cmss10 scaled \magstep1

\begingroup
\parindent=20pt
\thispagestyle{empty}

\advance\SaveW by -6.5in
\divide\SaveW by 2
\advance\SaveW by -\hoffset
\advance\SaveH by -8.9in
\divide\SaveH by 2
\advance\SaveH by -\voffset

\vbox to 0pt{
\vskip-\SaveH
\moveleft-\SaveW\vbox to 8.9in{\hsize=6.5in
\centerline{\twelvess BROOKHAVEN NATIONAL LABORATORY}
\vskip6pt
\hrule
\vskip1pt
\hrule
\vskip4pt
\hbox to \hsize{November, 2002 \hfil BNL-HET-02/22}
\vskip3pt
\hrule
\vskip1pt
\hrule
\vskip3pt

\vskip1in
\centerline{\LARGE\bf SUSY Signatures at LHC}
\vskip.5in
\centerline{\bf Frank E. Paige}
\vskip4pt
\centerline{\it Physics Department}
\centerline{\it Brookhaven National Laboratory}
\centerline{\it Upton, NY 11973 USA}

\vskip.75in

\centerline{ABSTRACT}

\vskip8pt
\narrower\narrower

	The ATLAS and CMS Collaborations at the CERN Large Hadron
Collider (LHC) have devoted considerable effort to the study of SUSY
signatures and measurements. This talk provides an overview of what
can be learned at the LHC if TeV-scale SUSY exists.

\vskip1in

        Invited talk at the {\sl SUSY02: The 10th International
Conference on Supersymmetry and Unification of Fundamental
Interactions} (June 17--23, 2002, DESY, Hamburg).

\vskip0pt

\vfil\footnotesize
        This manuscript has been authored under contract number
DE-AC02-98CH10886 with the U.S. Department of Energy.  Accordingly,
the U.S.  Government retains a non-exclusive, royalty-free license to
publish or reproduce the published form of this contribution, or allow
others to do so, for U.S. Government purposes.

\vskip0pt} 
\vss} 

\newpage
\thispagestyle{empty}
\
\newpage
\endgroup
\setcounter{page}{1}

\centerline{\Large\bf SUSY Signatures at LHC}
\medskip
\centerline{\bf Frank E. Paige}
\centerline{\it Brookhaven National Laboratory, Upton, NY 11973 USA}

\begin{abstract}

	The ATLAS and CMS Collaborations at the CERN Large Hadron
Collider (LHC) have devoted considerable effort to the study of SUSY
signatures and measurements. This talk provides an overview of what
can be learned at the LHC if TeV-scale SUSY exists.

\end{abstract}

\section{Introduction}

	SUSY is perhaps the most promising candidate for physics
beyond the Standard Model; it also provides a good test of detector
performance. The ATLAS~\cite{atlas} and CMS~\cite{cms} Collaborations at
the CERN Large Hadron Collider (LHC) have therefore devoted a lot of
effort to studying SUSY signatures and measurements. This talk
provides an overview of that work with emphasis on new results since
the overviews in Ref.~\citenum{AtlasTDR} and \citenum{Abdullin:1998pm}.

	The SUSY cross section at the LHC is dominated by the
associated strong production of gluinos and squarks. If $R$ parity is
conserved, these decay into the lightest SUSY particle $\lsp$, which
escapes the detector, plus quarks, gluons, and perhaps other Standard
Model particles. Thus, SUSY provides signatures containing at least
jets and large missing transverse energy $\etmiss$. The LHC should be
able to observe these signals for $\tg$ and $\tq$ masses up to about
$2\,\TeV$ with only $10\,\fbi$ of integrated luminosity.

	The challenge at the LHC is not to discover TeV-scale SUSY
(assuming it exists) but to make precision measurements of masses and
other quantities. Since the decay products of each SUSY particle contain
an invisible $\lsp$, no mass peaks can be reconstructed directly.
Instead, masses must be inferred~\cite{Hinchliffe:1997iu} from kinematic
endpoints and other properties of the events.  Developing methods to do
this has been a main emphasis of the studies to date. Typically, events
are simulated for a particular SUSY model and for the Standard Model
backgrounds using a parton shower Monte Carlo program such as
HERWIG~\cite{Corcella:2001wc}, ISAJET~\cite{Baer:1999sp}, or
PYTHIA~\cite{Sjostrand:2001yu}, the detector response is simulated using
a fast parameterization, cuts are made to give a good signal/background,
and various kinematic distributions are reconstructed. A number of
examples are presented below.

\section{Search for SUSY} 

	Since $\tg$ and $\tq$ are strongly produced, their cross
sections are comparable to QCD at the same $Q^2$. If $R$ parity is
conserved, their decays produce distinctive events with large
$\etmiss$. A typical analysis requires at least four jets with
$E_T>100,50,50,50\,\GeV$ and $\etmiss>100\,\GeV$ and plots as a measure
of $Q^2$ the quantity
$$
\Meff = \etmiss + \sum_{{\rm jets}\ j} p_{Tj}\,.
$$
For large $\Meff$ the Standard Model background is typically 10\% of
the signal.

\begin{figure}[t]
\dofig{4.5in}{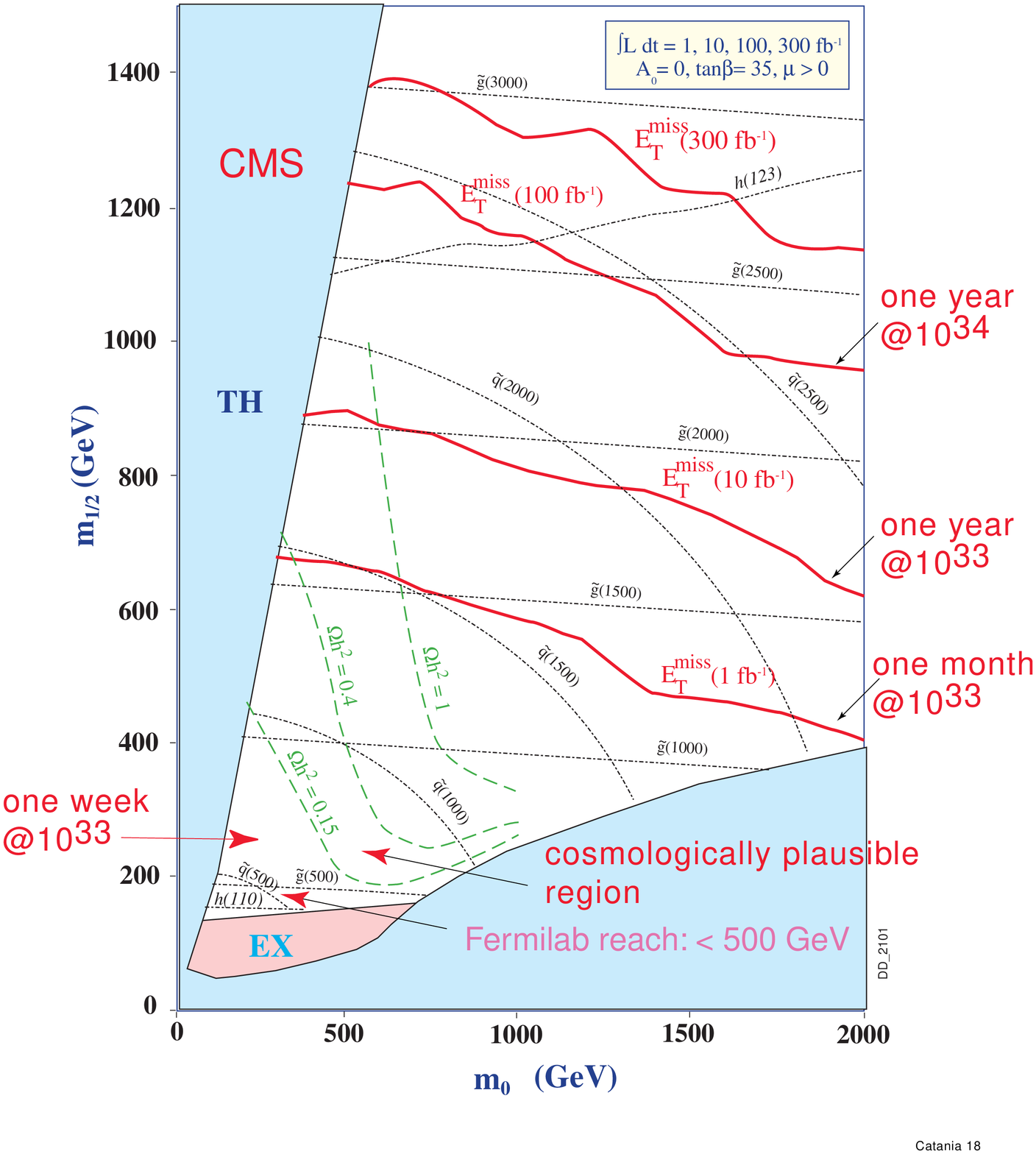}
\caption{Plot of $5\sigma$ reach in ${\rm jets} + \etmiss$ channel for
mSUGRA model~\cite{Abdullin:1998pm,Denegri}. 
\label{Catania18}}
\end{figure}

\begin{figure}[t]
\dofig{3.5in}{D_Denegri_1266c.mod}
\caption{Reach limits in various channels for
$100\,\fbi$~\cite{Abdullin:1998pm}. \label{D_Denegri_1266c}}
\end{figure}

	The reach is thus limited mainly by the $\tg$ and $\tq$
production cross sections and hence masses. This reach is shown for
the minimal SUGRA model in Fig.~\ref{Catania18}. Note that the
sensitivity for one month at 10\% of design luminosity is about
$1500\,\GeV$. Of course it will take more than one month to understand
the detectors, but nevertheless one might hope for an interesting talk
at SUSY08. Squark and gluino decays typically involve one or more
intermediate charginos or neutralinos, giving rise to multi-lepton
signatures as well as ones involving jets. The reach for such
signatures is shown in Figure~\ref{D_Denegri_1266c}. 

\begin{figure}[t]
\dofigs{2.95in}{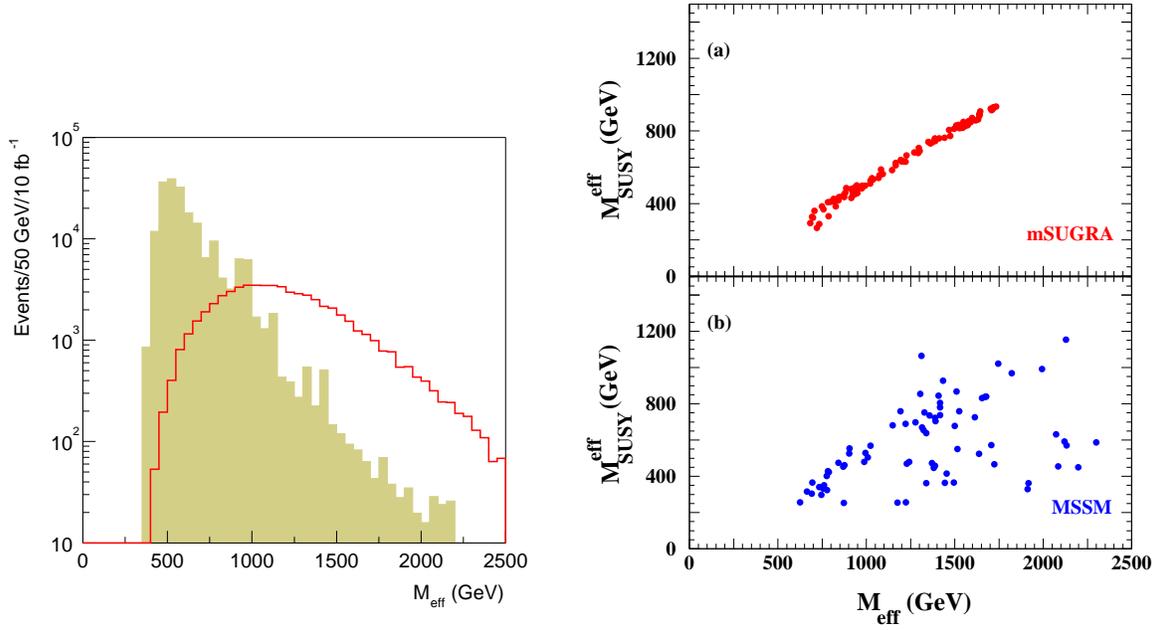}{tovey02.epsi}
\caption{Left: Typical $\Meff$ distribution. Right: Correlation of
$\Meff$ with SUSY mass scale~\cite{Tovey:2000wk}. \label{tovey02}}
\end{figure}

	The emergence of the SUSY signal from the Standard Model
background is illustrated for a typical case in Figure~\ref{tovey02}.
Also shown is the correlation between the peak of the $\Meff$
distribution and 
$$
M_{\rm SUSY} \equiv {\sum\nolimits_i M_i \sigma_i \over 
\sum\nolimits_i \sigma_i}, \qquad
M_{\rm SUSY}^{\rm eff} \equiv 
M_{\rm SUSY} - {M^2(\lsp) \over M_{\rm SUSY}}
$$
Obviously, however, any such inclusive correlation is model dependent.

	While the reach plots in Figure~\ref{Catania18} apply only to
mSUGRA, similar reach in gluino and squark masses should apply to any
model in which they decay into an invisible and relatively light
$\lsp$. Some models are easier. For example, GMSB models with prompt
$\lsp \to \tG \gamma$ or $\tell \to \tG \ell$ decays provide
additional photon or lepton handles to suppress Standard Model
backgrounds. GMSB models with a quasi-stable $\tell$ provide a
penetrating charged particle with $\beta<1$ that is well measured by
the muon detectors. AMSB models give few single leptons but larger
$\etmiss$. If $R$ parity is violated via $\lsp \to \ell^+\ell^-\nu$ or
$q \bar q \ell, q \bar q \nu$, there are again additional leptons.
Perhaps the most difficult case is $R$-violation via $\lsp \to cds$,
giving multiple jets but no $b$ jets. The background is poorly known,
but it seems likely that one must rely on leptonic cascade decays. In
general, however, discovery of gluinos and squarks with masses of
order $1\,\TeV$ appears straightforward. Masses of $3\,\TeV$ are of
course much more difficult and would probably need a luminosity
upgrade.

\section{\bf SUSY Measurements using Leptons} 

	If $R$ parity is conserved, then the $\lsp$ is invisible and
there are no mass peaks. Nevertheless, kinematic endpoints allow one
to determine mass combinations. The simplest example is the decay

\begin{figure}[t]
\dofig{4in}{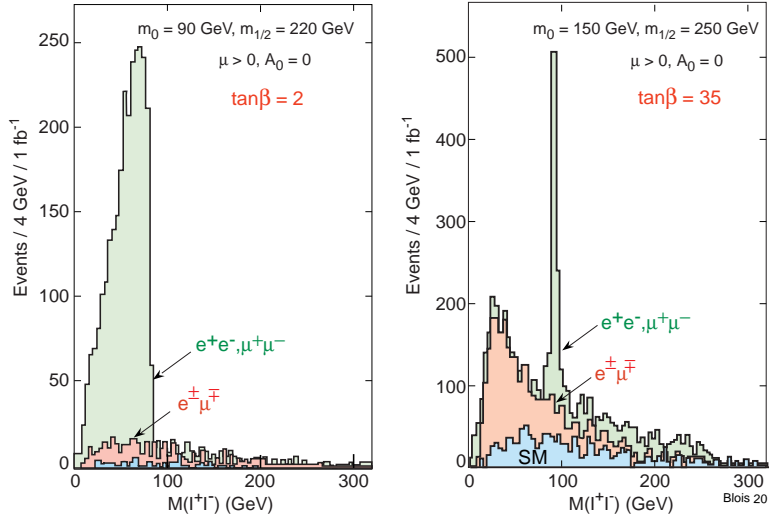}
\caption{Plot of $e^+e^-$, $\mu^+\mu^-$  and  $e^\pm\mu^\mp$ mass
distributions for mSUGRA with 2-body cascade decay (left) and $Z$ plus
3-body decay (right)~\cite{bloisref}.\label{blois}}
\end{figure}

\begin{figure}[t]
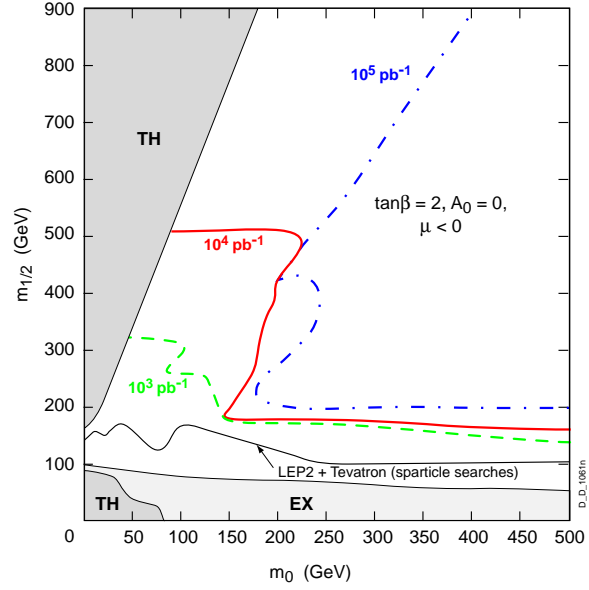

\dofig{3in}{cmsll.epsi}
\caption{Reach for observing dilepton endpoints in SUGRA models with
$1\,\fbi$, $10\,\fbi$ and $100\,\fbi$. Theory (TH) and experimental
constraints are also indicated~\cite{Abdullin:1998pm}.
\label{cmsll}}
\end{figure}

	For the dilepton signature, events are selected to have two
leptons with (typically) $p_{T,\ell}>10\,\GeV$ and $|\eta_\ell|<2.5$
in addition to multiple hard jets and large $\etmiss$. Then the
dominant Standard Model background is $t \bar t$. The background from
$t \bar t$ or from any other Standard Model source involving two
independent decays cancels in the combination
$e^+e^- + \mu^+\mu^- - e^\pm\mu^\mp$. The same is true for two
independent chargino decays in mSUGRA, and it is likely to be true in
any SUSY model that avoids $\mu \to e\gamma$ and other low-energy
constraints. 

\begin{figure}[t]
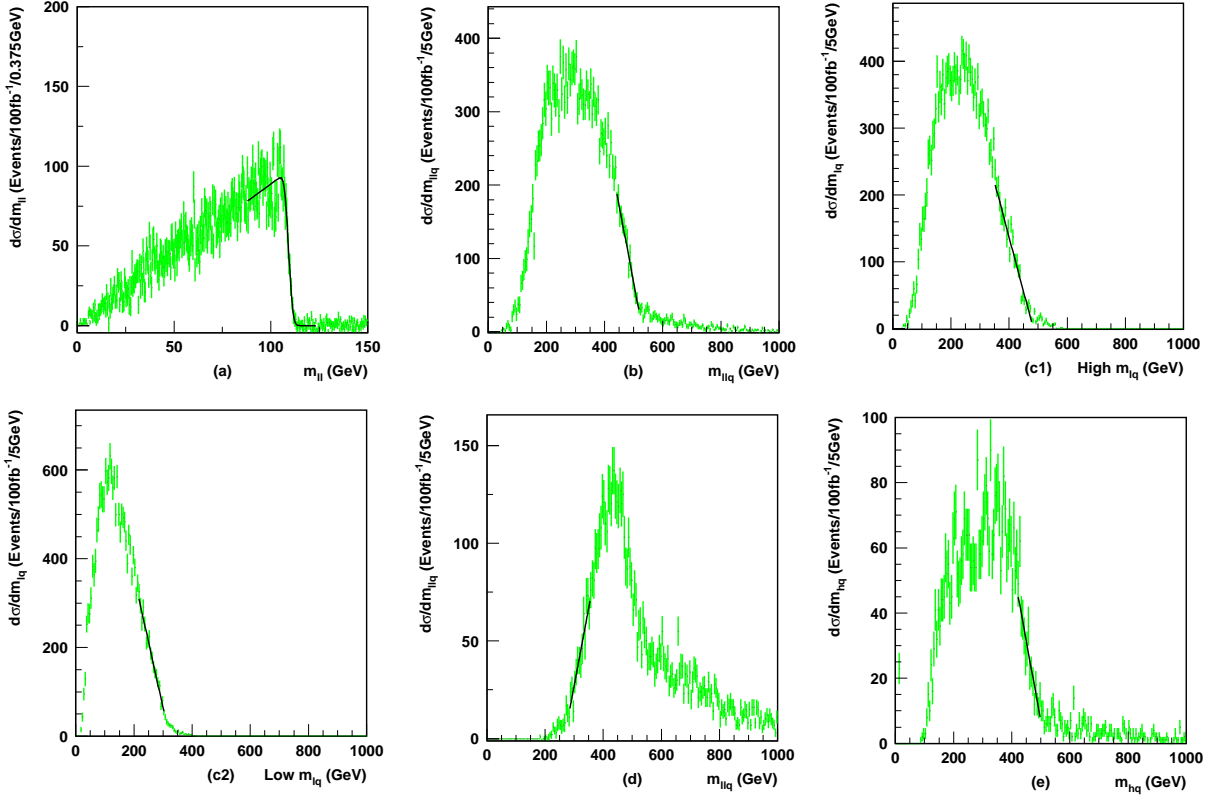

\dofig{\textwidth}{camb5new.epsi}
\caption{Dilepton + jet distributions for mSUGRA Point 5 as described
in the text. \label{camb5new}}
\end{figure}

\begin{figure}[t]
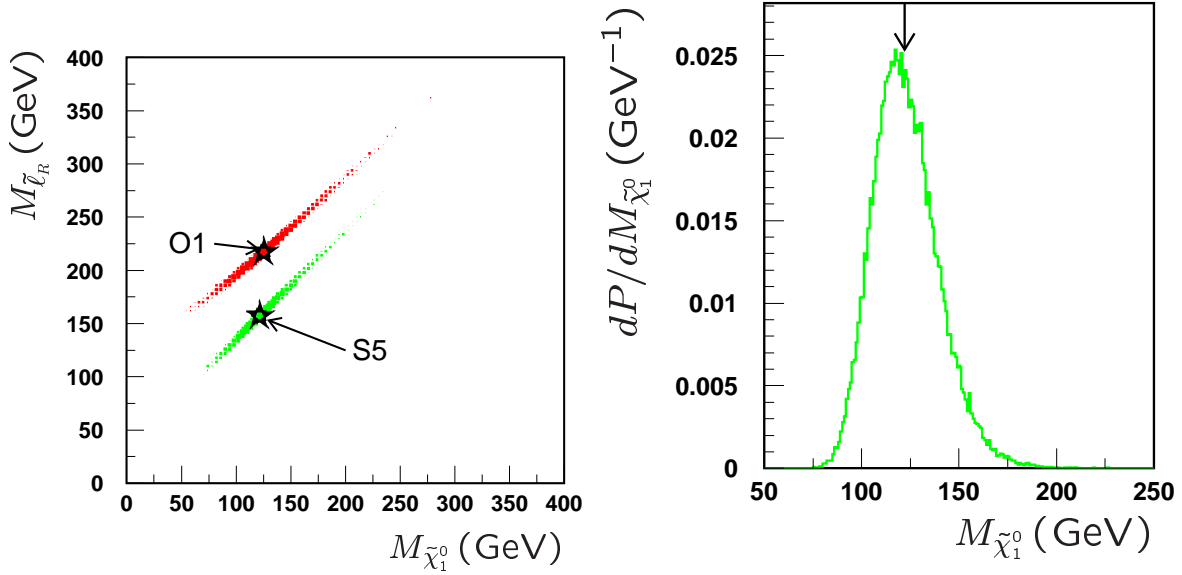

\dofigs{3in}{camb15new.epsi}{camb11new.epsi}
\caption{Left: Scatter plot of reconstructed values of $m_\ell \equiv
M_{\tell_R}$ vs.{} $m_1 \equiv M_{\lsp}$ for LHC Point~5 (S5) and
for an ``optimized string model'' (O1) using multiple measurements
from the decay chain $\tq_L \to \tchi_2^0 q \to\tell_R^\pm \ell^\mp q
\to \lsp\ell^+\ell^-q$. The stars mark the input values. Right:
Projection onto $M_{\lsp}$ axis~\cite{Allanach:2000kt}. \label{camb15}}
\end{figure}

	An opposite-sign, same-flavor dilepton signature can arise
from a 2-body cascade decay $\tchi_2^0 \to \tell^\pm \ell^\mp \to \lsp
\ell^+\ell^-$, from a 3-body $\tchi_2^0 \to \lsp \ell^+\ell^-$ decay,
or from a decay via a $Z$. All three cases are illustrated in
Figure~\ref{blois}. They are clearly distinguished by shape. Such
endpoints are observable over a significant fraction of the mSUGRA
parameter space, as illustrated in Figure~\ref{cmsll}. In particular,
a large part of the mSUGRA parameter space with acceptable cold dark
matter has light sleptons and hence enhanced $\ell^+\ell^-$ decays. 

	When a longer decay chain can be identified, more combinations
of masses can be measured. For example, at mSUGRA Point~5, the
dominant source of dileptons is $\tq_L \to \tchi_2^0 q \to \tell_R^\pm
\ell^\mp q \to \lsp\ell^+\ell^-q$. The gluino is heavier than the
squarks, and the hardest jets generally come from the squarks. In such
a case, one can form an $\ell^+\ell^- q$ endpoint and two $\ell^\pm q$
endpoints in addition to the $\ell^+\ell^-$
one~\cite{AtlasTDR,Bachacou:2000zb}. In addition, if a lower limit
on the $\ell^+\ell^-$ mass is imposed, there is also an $\ell^+\ell^-
q$ threshold. 

	All of these distributions after experimental selections are
shown in Figure~\ref{camb5new}. The endpoints and thresholds can all
be expressed in terms of the masses involved using elementary
kinematics, and they provide enough constraints to determine the
masses. The results taking estimated errors into account for the
$\lsp$ and $\tell_R$ masses are shown in Figure~\ref{camb15} for two
models with similar masses. All the masses involved are determined
quite accurately as functions of $M_\lsp$; in particular the two
models can be easily distinguished. The $\lsp$ mass is determined to
about 10\% through its effect on the kinematics even though it is
invisible.

\begin{figure}[t]
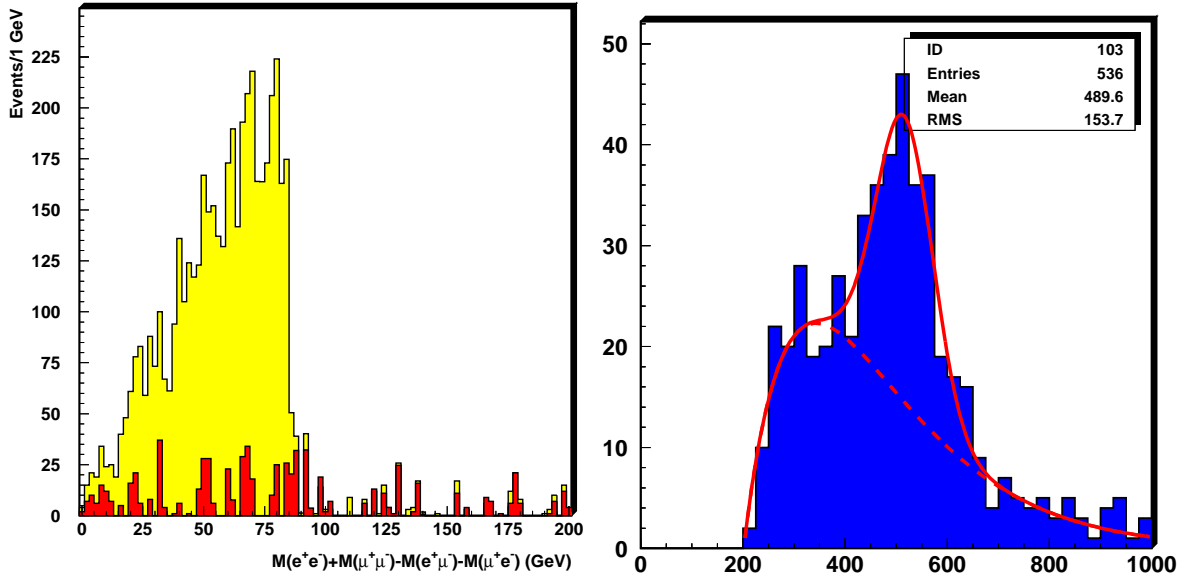

\dofigs{3in}{mll_etmiss_50.epsi}{sb_fit_nocut_b_10fb.epsi}
\caption{Left: Dilepton distribution for Point B analysis. Right:
$M(\tchi_2^0 b)$ distribution using inferred $\tchi_2^0$
momentum\cite{Tricomi}. \label{Tricomi}}
\end{figure}

	An alternative approach is to reconstruct $\tchi_2^0$ momentum
assuming that the $\lsp$ mass is known and using
$$
\vec p_{\tchi_2^0} = \vec p_{\ell\ell} \left(1 + {M_\lsp \over
M_{\ell\ell}} \right)
$$
While this is exact only for a 3-body $\tchi_2^0 \to \lsp \ell\ell$ at
the endpoint, it can be used as an approximation elsewhere. An
analysis for SUGRA Point B ($m_0=100\,\GeV$, $\mhalf=250\,\GeV$,
$A_0=0$, $\tan\beta=10$, $\mu>0$) using this method looks for $\tg \to
\tilde b \bar b \to \tchi_2^0 b \bar b \to \tell \ell b \bar b \to
\lsp \ell\ell b \bar b$.  The dilepton distribution after cuts and the
reconstructed $\tchi_2^0 b$ distribution using
$75<M_{\ell\ell}<92\,\GeV$ are shown in Figure~\ref{Tricomi}. A second
$b$ can be added to reconstruct the gluino mass; see
Ref.~\citenum{Tricomi}.

\section{\bf\boldmath SUSY Measurements using $h \to b \bar b$} 

\begin{figure}[t]
\dofig{4in}{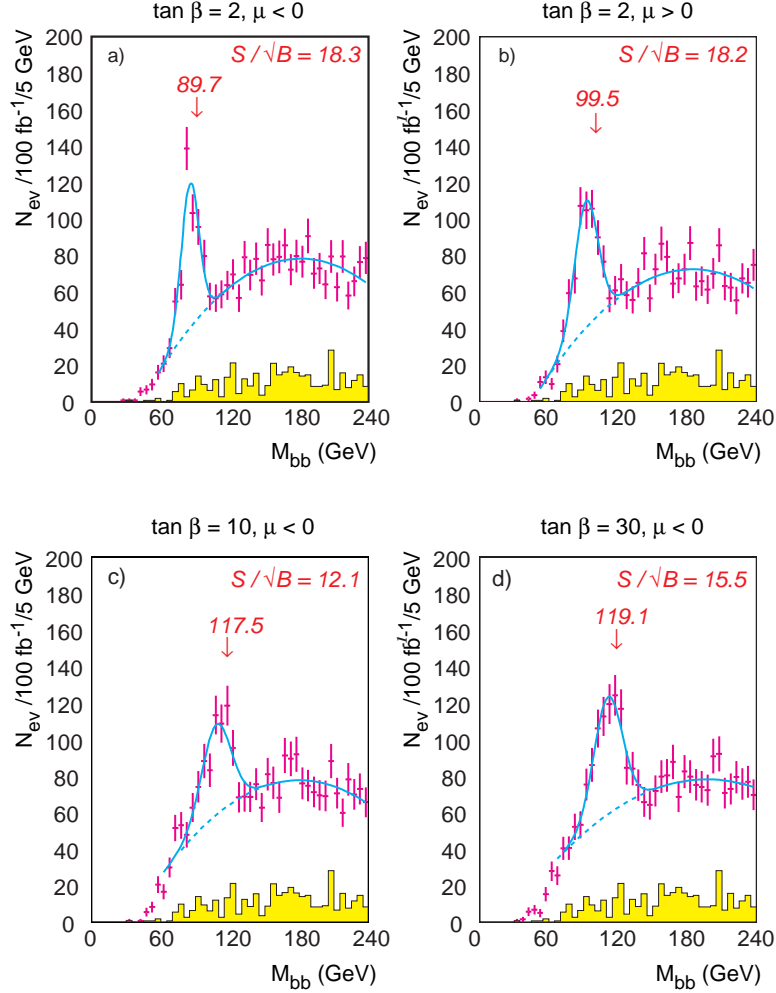}
\caption{Plot of $b \bar b$ dijet mass distribution (points) with $h
\to b \bar b$ signal (solid), SUSY background (dashed), and Standard
Model background (shaded) for various
$\tan\beta$\cite{cmssusyb,Abdullin:1998pm}. \label{cmshbb}}
\end{figure}

	If the decay $\tchi_2^0 \to \lsp h$ is kinematically allowed,
it often has a substantial branching ratio. The decay $h \to b \bar b$
can be reconstructed by selecting events with multiple jets plus large
$\etmiss$ and plotting the mass of pairs of jets tagged as $b$'s. The
reach for this signature in mSUGRA, shown in Figure~\ref{cmshbb},
covers a substantial fraction of parameter space. It might well be the
discovery mode for the light SUSY Higgs.

\begin{figure}[t]
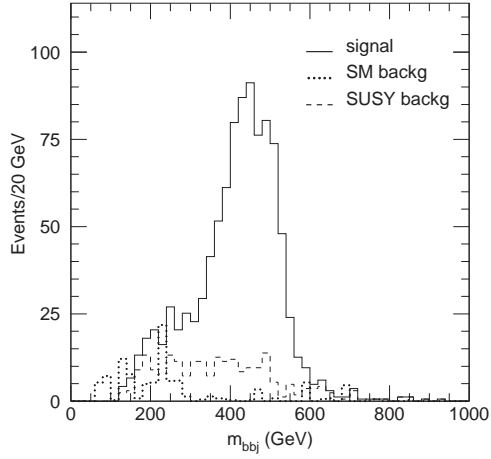

\dofig{2.5in}{p5_mbbj.epsi}
\caption{Minimum $M(hj)$ mass from $h \to b\bar b$ combined with two
hardest jets~\cite{AtlasTDR}. \label{p5_mbbj}}
\end{figure}

	If $h \to b \bar b$ can be reconstructed, it can then be
combined with jets to determine other masses.  In Figure~\ref{p5_mbbj}
for example, the $h$ was combined with each of the two hardest jets in
the event; the smaller of these masses should be less than the $\tq \to
\lsp h q$ endpoint. This measurement is less precise than those
involving leptons, but it may be useful if the leptonic branching
ratios are small.

\section{Complex SUSY Signatures}

	SUSY events can be much more complex than those for the cases
discussed above. An example is the ``focus point'' region: in mSUGRA as
$m_0$ becomes large for fixed values of the other parameters, $\mu\to0$
and radiative electroweak symmetry breaking fails. Near this boundary,
the $\lsp$ is dominantly Higgsino, leading to acceptable values of cold
dark matter~\cite{Feng:2000gh}. The exact location of the focus point
region is very sensitive to details of the SUSY spectrum
calculation~\cite{Allanach:2002pz}, but such a region surely exists. 

\begin{figure}[t]
\dofig{3in}{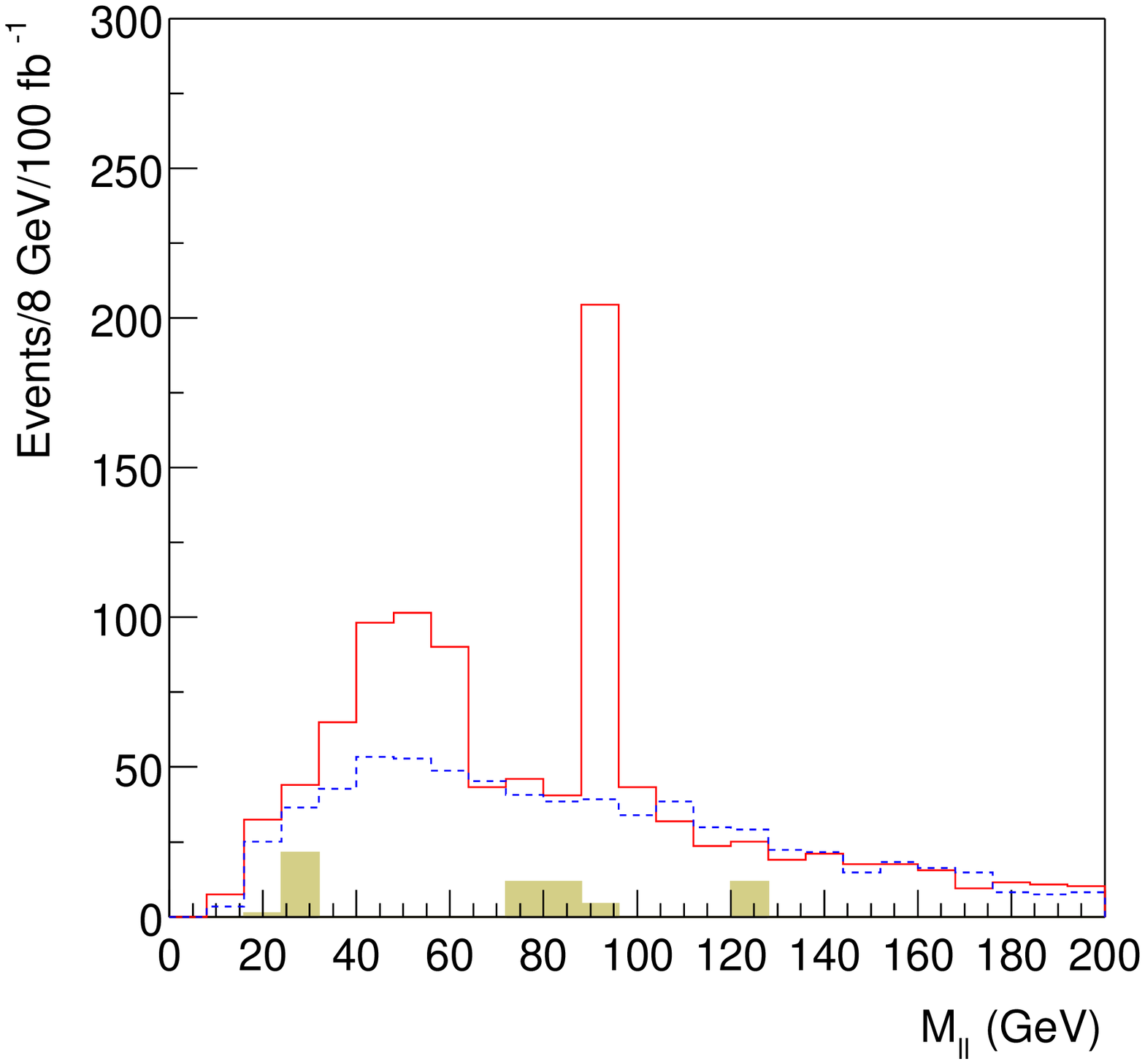}
\caption{Dilepton mass spectrum for focus point SUSY
model\cite{Hinchliffe:2001bz,Hinchliffe:2001by}. \label{c1500mll}}
\end{figure}

	In one calculation~\cite{Battaglia:2001jn} the mSUGRA point
$m_0=1500\,\GeV$, $\mhalf=300\,\GeV$, $A_0=0$, $\tan\beta=10$, $\mu>0$
is close to the $\mu=0$ focus point boundary. Since the squarks are
heavy, SUSY production is dominated by $\tg\tg$ pairs. The largest $\tg$
branching ratios are calculated~\cite{Baer:1999sp} to be $B(\tg
\to\tchi_1^- t \bar b + \hc) \approx B(\tg \to \tchi_2^- t \bar b + \hc)
\approx 23\%$. These decays give events with 12--16 jets and leptons!
The methods described previously suffice to find such signals. Many of
the basic starting points discussed previously also work; see for
example Figure~\ref{c1500mll}. But sorting out the combinatorial
background in such complex events is quite difficult and has not yet
been solved.

	$R$-parity violation with $\lsp \to qqq$, and especially $\lsp
\to cds$ is another example of a SUSY model with very complex
signatures. In such models the signal has very high jet multiplicity ---
nominally six jets just from the $\lsp$ decays --- few or no $b$ jets
that might be used to reduce the combinatorial background, and small
$\etmiss$. The QCD background for such events is not well known; the
1-loop QCD corrections to the production of 6-10 jets seem unlikely to
be calculated even by the end of LHC running. Estimates based on QCD
shower Monte Carlo programs suggest that the backgrounds are comparable
to the signals, but this is at best a rough estimate.

\begin{figure}[t]
\dofig{3in}{camb6a.epsi}
\caption{Scatter plot of $M^+$ vs.\ $M^-$ for $R$-violating $\lsp \to
qqq$ decays\cite{Allanach:2001xz}. \label{camb6a}} 
\end{figure}

\begin{figure}[t]
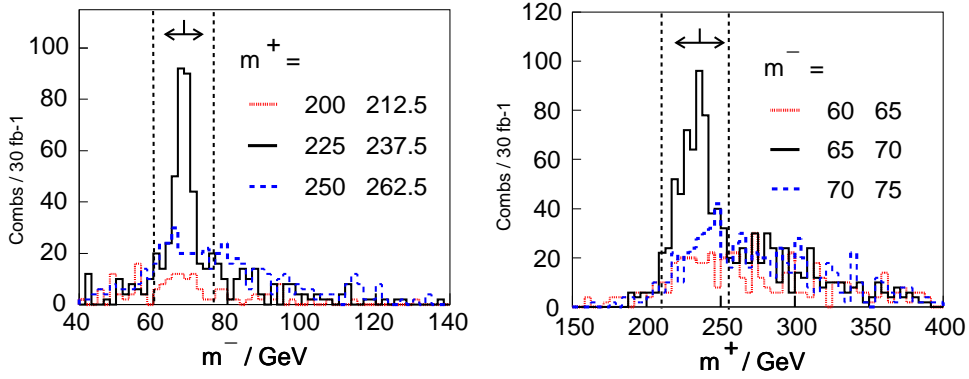

\dofig{5in}{camrviolnew.epsi}
\caption{Distributions of $m^\pm = M(\ell^+\ell^-qqq) \pm M(qqq)$ in
$R$-parity violating SUSY events with $\lsp \to qqq$ after selecting
either the peak or sidebands in $m^\mp$~\cite{Allanach:2001xz}.
\label{camrviol}}
\end{figure}

	For such scenarios, therefore, it is probably necessary to rely
on SUSY cascade decays, e.g., $\tchi_2^0 \to \tell^\pm\ell^\mp \to \lsp
\ell^+\ell^- \to qqq\ell^+\ell^-$, to provide additional handles to
suppress the QCD background. One example that has been studied uses such
decays with branching ratios taken from mSUGRA Point 5. Even though
there are no missing particles, reconstruction is difficult because the
jets from the $\lsp$ are soft and the mass resolution is only
$\sim10\%$. Define the mass combinations $M^\pm = M(qqq\ell^+\ell^-) \pm
M(qqq)$; then the jet resolution largely cancels for $M^-$, as can be
seen in Figure~\ref{camb6a}. The distributions in Figure~\ref{camrviol}
show reasonable distributions in these variables.

	These are just two examples of SUSY scenarios giving complex
signatures. Such scenarios are certainly possible, and much more work is
needed to develop analysis strategies to deal with them.

\section{\boldmath SUSY Signatures with $\tau$'s} 

	The initial motivation for studying $\tau$ signatures came
from the fact that for large regions of mSUGRA parameter space with
$\tan\beta\gg1$ the only two-body decay modes are $\tchi_2^0 \to
\ttau_1\tau$ and $\tchi_1^\pm \to \ttau_1^\pm \nu_\tau$. Then these
modes have branching ratios close to 100\%, and all SSUY decays involve
$\tau$'s. Such scenarios are disfavored because they tend to give
contributions to $g_\mu-2$ much larger than observed. 

	Low-energy tests of the Standard Model seem to require that the
first two generations of squarks and sleptons be nearly degenerate. This
degeneracy is not understood in general but does occur in mSUGRA. Even
in mSUGRA, however, the third generation of squarks and sleptons is
split by renormalization group and left-right mixing effects. Thus,
study of the third generation is likely to be essential for
understanding SUSY.

\begin{figure}[t]
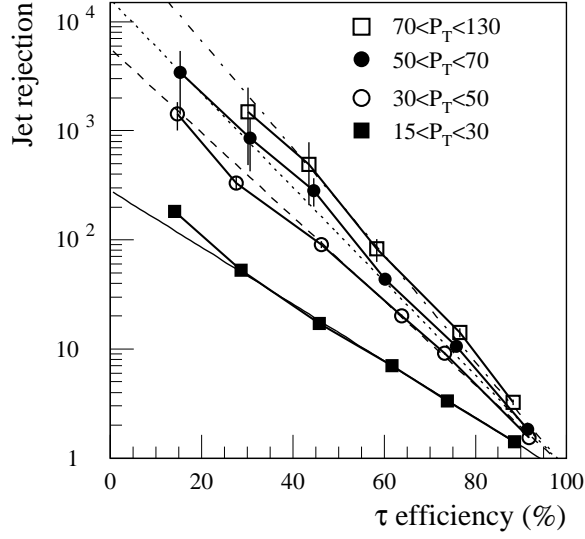

\dofig{3in}{tdr9-31.epsi}
\caption{QCD jet rejection vs.\ efficiency for hadronic $\tau$ decays
based on full simulation of the ATLAS detector~\cite{AtlasTDR}.
\label{tdr9-31}}
\end{figure}

	Because of technical constraints, the ATLAS and CMS vertex
detectors do not have sufficient resolution to identify $\tau$ decays
cleanly. Hence, leptonic $\tau$ decays cannot be distinguished from
prompt leptons, and $\tau$'s can only be identified using narrow,
one-prong jets, for which there is substantial QCD background. As can be
seen from Figure~\ref{tdr9-31}, the typical jet rejection is a factor of
about 100 for a $\tau$ efficiency of 50\%.

\begin{figure}[t]
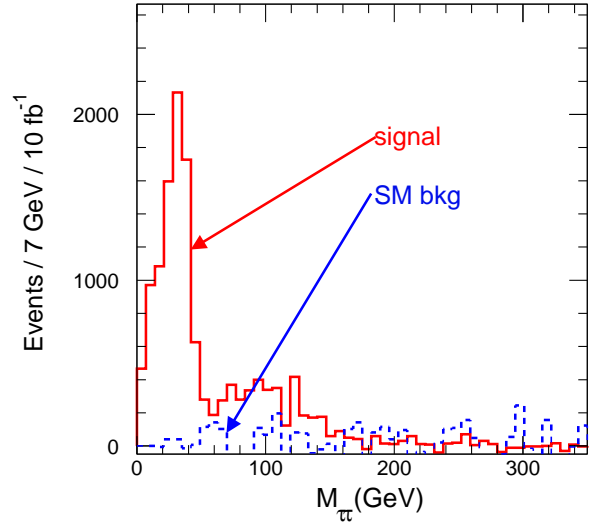

\dofig{3in}{p6subtracted.epsi}
\caption{Plot of $\tau^+\tau^- - \tau^\pm\tau^\pm$ visible mass
distribution with hadronic $\tau$ decays at LHC SUGRA
Point~6~\cite{AtlasTDR}. \label{p6subtracted}}
\end{figure}

	For mSUGRA with $m_0=\mhalf=200\,\GeV$, $A_0=0$,
$\tan\beta=45$, the only 2-body decays of light gauginos are
$\tchi_2^0 \to \ttau_1 \tau$ and $\tchi_1^\pm \to \ttau_1 \nu_\tau$.
These decays therefore dominate. Since there is $\etmiss$ both from
$\nu_\tau$'s and from $\lsp$'s, the $\tau$ distributions can only be
inferred from their hadronic decay products. Instead of a sharp
$\tau\tau$ edge at $59.64\,\GeV$, one would observe the distribution
shown in Figure~\ref{p6subtracted}. 

\begin{figure}[t]
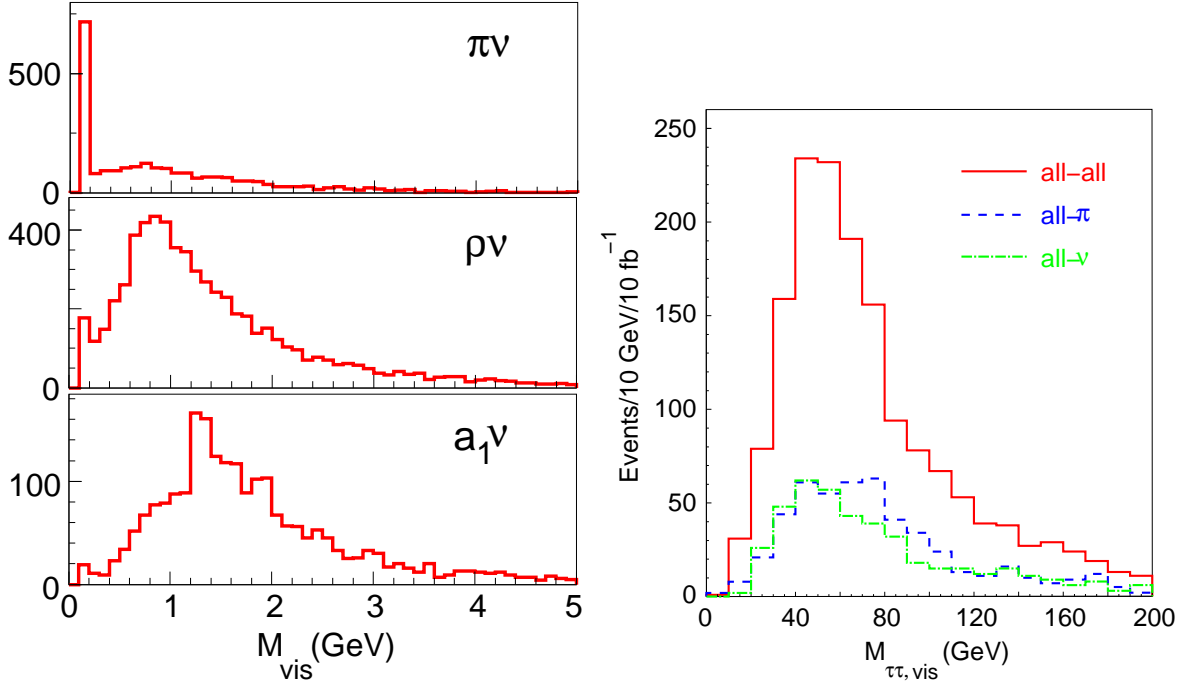

\dofigs{3in}{mtaunew.epsi}{lvlund22.epsi}
\caption{Left: Reconstructed masses for various $\tau$ decay modes.
Right: $\tau\tau$ mass distribution for all modes, for one
$\tau\to\pi\nu$ decay, and for decays with reversed helicities.
\label{mtaunew}}
\end{figure}

	The visible $\tau$ momentum depends both on the momentum and
on the polarization of the parent $\tau$. The decay $\tau \to \pi\nu$
is maximally sensitive to the polarization while high-mass decays are
rather insensitive to it. A study of the separation of different decay
modes has been carried out using a full GEANT-based simulation of
$\tau$'s from SUSY events in the ATLAS detector. The visible $\tau$
mass is reconstructed by combining charged tracks and electromagnetic
calorimeter cells, exploiting the fine granularity of the calorimeter. 
The reconstructed masses, Figure~\ref{mtaunew}, show a good separation
of $\pi\nu$ decays and some difference between $\rho\nu$ and $a_1\nu$
decays. The same figure indicates the sensitivity of the $\tau\tau$
mass distribution to a reversal of helicities. 

\section{GMSB Signatures}

	In GMSB models the lightest SUSY particle is the light
gravitino $\tG$. The phenomenology depends on nature and lifetime of
the next lightest particle (NLSP), which can be either the $\lsp$ or a
$\tell$. Decays into $\tG$ lead to more distinctive events and to
longer decay chains, such as $\tchi_2^0 \to \tell^\pm \ell^\mp \to
\lsp \ell^+\ell^- \to \tG \gamma\ell^+\ell^-$, that can be used to
determine SUSY masses~\cite{AtlasTDR}.

\begin{figure}[t]
\dofigs{3in}{tdr20-65.epsi}{tdr20-66.epsi}
\caption{Angle (left) and time delay (right) for photons from
long-lived $\lsp \to \tG \gamma$ in the ATLAS detector\cite{AtlasTDR}.
\label{tdr20-65}}
\end{figure}

\begin{figure}[t]
\dofigs{3in}{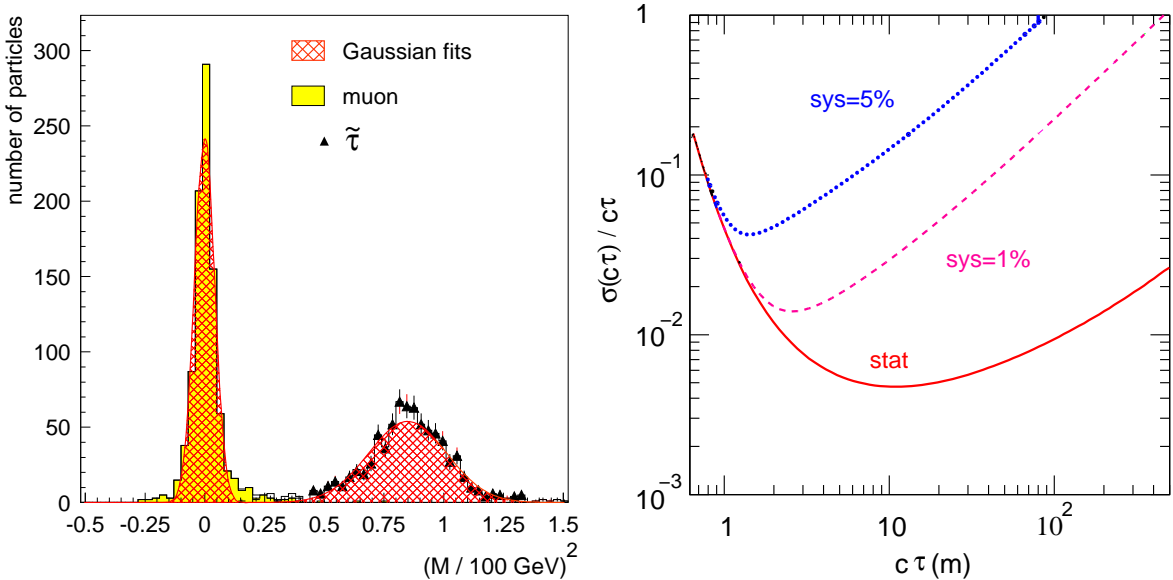}{staulife.epsi}
\caption{Left: Time of flight separation of muons and quasi-stable
$\tell$~\cite{cmsnlsp}. Right: Measurement of $\tell$ lifetime for
different estimates of the systematic error on the
acceptance~\cite{Ambrosanio:2000zu}. \label{cmsmstau}}
\end{figure}

	The lifetime for the NLSP to decay to a $\tG$ is unknown and
is an important parameter of the model. Short $\lsp \to \tG \gamma$
lifetimes can be measured by selecting Dalitz decays and using the
vertex detector. For $c\tau_\lsp \gg 1\,{\rm m}$ the signature is
occasional photons that do no point to the primary vertex. The ATLAS
electromagnetic calorimeter was designed to measure photon angles with
$\Delta \theta \approx {60\,{\rm mr} \over \sqrt{E}}$ to improve the
$h \to \gamma\gamma$ mass resolution. It also has good time
resolution, $\Delta t \approx 100\,{\rm ps}$. Both can be exploited to
detect long-lived $\lsp \to \tG \gamma$ decays with good efficiency,
as seen in Figure\~ref{tdr20-65}. If no signal is seen, an upper limit
of order $c\tau\sim100\,{\rm km}$ could be established~\cite{AtlasTDR}.

	A long-lived slepton NLSP would give tracks with $\beta<1$
through the calorimeter and muon system. These can be well identified
by using the muon system as a time of flight system, as can be seen in
Figure~\ref{cmsmstau}~\cite{AtlasTDR,cmsnlsp}. The lifetime can be
estimated by counting the number of events with zero, one, and two
such tracks, as is also shown in Figure~\ref{cmsmstau}. It should also
be possible to determine the lifetime by looking for kinks in the
central tracker, but developing the needed pattern recognition
algorithm is non-trivial and has not yet been attempted.

\section{Outlook}

	If SUSY exists with masses below about $1\,\TeV$, it seems
likely that the ATLAS and CMS detectors should find evidence for it
shortly after the LHC begins operation. Only a limited number of SUSY
models and cases have been investigated so far, and in all cases the
answer has been known. These studies seem sufficient, however, to
sketch the broad outlines of an initial program of SUSY measurements
(assuming that $R$ parity is conserved):

\begin{enumerate}

\item Search for an excess of multijet + $\etmiss$ events over
Standard Model predictions, which of course must be checked against
other measurements.

\item If such an excess is found, select a sample of SUSY with simple
cuts such as those described above.

\item Look for special features such as $\gamma$'s or long-lived
$\tell$'s in these events.

\item Look for $\ell^\pm$, $\ell^+\ell^-$, $\ell^\pm\ell^\pm$, $b$
jets, hadronic $\tau$ decays, etc.

\item Try simple endpoint-type analyses such as those described above.

\end{enumerate}

\noindent This program looks quite feasible. 

	There is of course much more information available, including
cross sections, branching ratios, and other kinematic distributions.
Ultimately one will want to use the first measurements to guide
further analyses incorporating all the information. This program will
need the full power of the LHC and its detectors.

\end{document}